\documentclass[aps,preprint]{revtex4}%
\usepackage{amsfonts}
\usepackage{amsmath}
\usepackage{amssymb}
\usepackage{graphicx}%
\begin{document}
\preprint{ }
\title{Consideration of the relationship between Kepler and cyclotron dynamics
leading to prediction of a non-MHD gravity-driven Hamiltonian dynamo}
\author{P. M. Bellan}
\affiliation{Applied Physics, Caltech, Pasadena CA 91125}
\pacs{PACS number}

\begin{abstract}
Conservation of \textit{canonical} angular momentum shows that charged
particles are typically constrained to stay within a poloidal Larmor radius of
a poloidal magnetic flux surface. However, more detailed consideration shows
that particles with a critical charge to mass ratio can have zero canonical
angular momentum and so be both immune from centrifugal force and not
constrained to stay in the vicinity of a specific flux surface. Suitably
charged dust grains can have zero canonical angular momentum and in the
presence of a gravitational field will spiral inwards \textit{across} poloidal
magnetic surfaces toward the central object and accumulate. This accumulation
results in a gravitationally-driven dynamo, i.e., a mechanism for converting
gravitational potential energy into a battery-like electric power source.

\end{abstract}
\volumeyear{ }
\volumenumber{ }
\issuenumber{ }
\eid{ }
\date{}
\maketitle

\section{Introduction}

To the best of the author's knowledge, no plasma has ever been observed to
make a Kepler orbit around the Sun, a planet, or a moon. For example, the
Earth's magnetotail does not make Kepler orbits around the Earth nor does the
solar wind make a Kepler orbit around the Sun. This is puzzling because in the
astrophysical literature plasmas are routinely presumed to make near-Kepler
orbits in the presence of the gravitational field of a central object, e.g.,
accretion disks are modeled as MHD\ plasmas in a near-Kepler orbit around a
central object. The qualifier `near' is used because it is conventionally
presumed for gases and plasmas that a radial gradient \ of an isotropic
pressure exists and that this pressure gradient provides a modest outward
force which slightly reduces the amount of centrifugal force required to
balance the inward force of gravity and so achieve a stable circular orbit
(e.g., see Ref. \cite{Weidenschilling:2000} for a discussion of the
implications of this effect in the context of a system consisting of gas and
solid particles).

If Kepler orbiting plasmas are so ubiquitous in astrophysics, then why is
there not even a single example in the great variety of plasma/gravitational
situations in our own solar system? A Kepler orbit is essentially a property
of a single point particle -- it is not a property of a collection of
independent point particles. For example, Earth and Mars are individually in
Kepler orbits around the Sun, but the center of mass of Earth and Mars is not
in a Kepler orbit around the Sun. If one replaced Earth and Mars by some
statistically large number of point particles, then each could be in its own
Kepler orbit around the Sun, but the center of mass of this configuration
would not be in a Kepler orbit.

A collisional gas in a container could be in a Kepler orbit around the Sun
because the walls of the container \textit{bind} the particles to stay within
a fixed distance of the center of mass so that the entire system can be
considered as a point particle located at the center of mass. The transition
from single point particle behavior to the behavior of a group of particles
can be seen by considering a moon in close orbit around a massive planet. The
tidal forces resulting from the gradient of the planet's gravitational force
can be so strong as to overcome the binding forces and fracture the moon. The
fragments would then follow distinct individual Kepler orbits and the center
of mass of these orbits would not follow a Kepler orbit. Similarly, in order
for a gas (or plasma) to have its center of mass follow a Kepler orbit there
would have to be some binding force, such as would be provided by container
walls, that would prevent the particle constituents of the gas (or plasma)
from separating onto distinct Kepler orbits. Collisions might provide binding
at the interior of a gas cloud, but not at the periphery because particles at
the periphery moving away from the center would not encounter other particles
with which they could collide. The periphery would simply expand into vacuum
if there is no wall to prevent this expansion and the particles constituting
the periphery would make Kepler orbits substantially different from the Kepler
orbit calculated for the center of mass.

\bigskip

Accretion disks are composed of dust and gas and the dust to gas mass ratio is
estimated \cite{Greaves:2000,Augereau:1999} to range from $\sim10^{-2}$ to
$\sim1.$ UV radiation photo-ionizes \cite{Wolf:2003,Lee:1996,Sickafoose:2000}
the dust and gas so the accretion disk can be considered as a dusty plasma
\cite{Verheest:2000} consisting of charged dust grains, electrons, and ions.
The magnetorotational instability (MRI) \cite{BalbusHawley:1991} and the
unipolar induction dynamo (UID) \cite{Lovelace:1976} assume accretion disks
are axisymmetric ideal MHD\ plasmas and neglect dusty plasma physics effects.
The MRI\ and UID\ additionally assume that accretion disks obey both Kepler
dynamics and ideal MHD. Thus accretion disks are considered to be ideal
MHD\ plasmas in a circular Kepler orbit about a central object; they are
supposed to conserve angular momentum and have frozen-in magnetic flux. One
can then ask what does the trajectory of an individual particle in the
accretion disk look like. Since MHD is assumed and MHD\ is based on the
assumption that all particles make cyclotron orbits, it seems that the
individual particle in question should be making a cyclotron orbit. On the
other hand since the whole plasma is supposed to be making a circular Kepler
orbit around the central object, then presumably this cyclotron-orbiting
particle must also be making a Kepler orbit around the central object. A
Kepler-orbiting cyclotron orbit is not an obvious concept to visualize, at
least to this author. This conceptual difficulty suggests that instead of
assuming that a particle is simultaneously Kepler-orbiting and cyclotron
orbiting, one should go back to first principles to investigate how Kepler and
cyclotron orbits relate to each other. Perhaps it will then become obvious how
to visualize a Kepler-cyclotron orbiting particle, or perhaps not.

\section{Kepler effective potential and orbits}

Let us begin with a brief review of Kepler orbits. A particle $\sigma$ with
conserved angular momentum $L_{\sigma}=m_{\sigma}r^{2}d\phi/dt$ has radial
motion in the Kepler \textquotedblleft effective\textquotedblright\ potential
\cite{Landau:1960}%
\begin{equation}
\chi_{Kepler}(r)=\frac{L_{\sigma}^{2}}{2m_{\sigma}r^{2}}\ -\frac{m_{\sigma}%
MG}{r} \label{Kepler effective potential}%
\end{equation}
where $M$ is the mass of a central object. Particles with energies at
$\min\chi_{Kepler}(r)$ have circular trajectories with $r=L_{\sigma}%
^{2}/m_{\sigma}^{2}MG$, velocity $v_{K}=\sqrt{MG/r}$ and constant angular
velocity $\Omega=\sqrt{MG/r^{3}}$ whereas particles with energy exceeding
$\min\chi_{Kepler}(r)$ have elliptical trajectories and variable angular
velocity $d\phi/dt=L_{\sigma}/m_{\sigma}r^{2}$ \cite{Landau:1960}.

\section{Guiding center orbits in combined magnetic and gravitational fields}

In contrast to Kepler dynamics \cite{Landau:1960}, basic plasma theory
\cite{Alfven:1950,Spitzer:1956,Sturrock:1994} shows that the guiding center of
\ a\ charged particle in combined magnetic and gravitational fields, but
\textit{no} electric field, \ drifts at the velocity%
\begin{equation}
\mathbf{v}_{g\sigma}=\frac{m_{\sigma}}{q_{\sigma}B^{2}}\mathbf{g\times
B}\ \ \label{vg}%
\end{equation}
where $\mathbf{g}$ is the gravitational acceleration. For a mass $M$ central
object, $\mathbf{g}=MG\nabla\left(  r^{2}+z^{2}\right)  ^{-1/2}\ $and so the
guiding center drift in the $z=0$ plane is
\begin{equation}
\mathbf{v}_{g\sigma}=\ \frac{MG}{r^{2}\omega_{c\sigma}}\hat{\phi}%
\ =\frac{\Omega}{\omega_{c\sigma}}v_{K}\hat{\phi} \label{vg 2nd way}%
\end{equation}
where $\omega_{c\sigma}=q_{\sigma}B_{z}/m_{\sigma}$ is the cyclotron
frequency.\ It is seen that $\mathbf{v}_{g}$ is \textit{ smaller} than the
Kepler velocity $v_{K}$ by $\Omega/\left\vert \omega_{c\sigma}\right\vert ,$
an enormous ratio for electrons and ions since $\left\vert \omega
_{ce}\right\vert \ $and $\left\vert \omega_{ci}\right\vert $ are many orders
of magnitude larger than $\Omega\ $for typical field strengths. If all the
particles move at a much slower velocity than the Kepler velocity, then how
could the center of mass move at the Kepler velocity? Furthermore, the
$\mathbf{v}_{g\sigma}$ add up to give the azimuthal current $\mathbf{J}%
_{g}=\sum n_{\sigma}q_{\sigma}\mathbf{v}_{g\sigma}=\rho\mathbf{g\times
B/}B^{2}$ where $\rho=\sum n_{\sigma}m_{\sigma}$. The gravitational drift
current results mainly from heavy particle motion \cite{Sturrock:1994} and
gravity is \textit{ }balanced\textit{ } by the magnetic force (i.e.,
$\mathbf{J}_{g}\times\mathbf{B=}-\rho\mathbf{g}$ \cite{Sturrock:1994}%
)$\mathbf{\ }$ rather than by centrifugal force which is insignificant for
this example. The ideal MHD\ Ohm's law $\mathbf{E+U\times B}=0$ actually fails
here, because the Hall term $\mathbf{J\times B}/ne$ in the zero-pressure,
generalized \cite{Goosens:2003} Ohm's law $\mathbf{E+U\times B=J\times B}/ne$
cannot be dropped since $\mathbf{J/}ne$ nearly equals the center of mass
velocity $\mathbf{U}.$

What sort of trajectory does an actual charged particle follow in an
astrophysical situation? Is it the Kepler orbit assumed in Refs.
\cite{Lovelace:1976,BalbusHawley:1991} or the much slower gravitational drift
derived in Ref. \cite{Sturrock:1994}? We show here that even though charged
particles in a strong magnetic field can rotate at $v_{K}$ as assumed in Refs.
\cite{Lovelace:1976,BalbusHawley:1991}, the motion is not governed by
Eq.\ref{Kepler effective potential} so charged particles (and hence a
plasma)\ do not in general obey Kepler dynamics. This analysis leads to the
realization that dust grains having a critical charge to mass ratio spiral in
towards the central object and so could provide a gravitationally powered
dynamo suitable for driving astrophysical jets.

\section{Hamiltonian description of orbits in combined electromagnetic and
gravitational fields}

We consider the axisymmetric charged particle Hamiltonian
\cite{Stormer:1955,Schmidt:1979,Lehnert:1971,Dullin:2002,Shebalin:2004}
\begin{align}
H  &  =\frac{m_{\sigma}v_{r}^{2}}{2}+\frac{m_{\sigma}v_{z}^{2}}{2}+\frac
{1}{2m_{\sigma}r^{2}}\left(  P_{\phi}-\dfrac{q_{\sigma}}{2\pi}\psi
(r,z,t)\right)  ^{2}\nonumber\\
&  +q_{\sigma}V(r,z,t)-\frac{m_{\sigma}MG}{\sqrt{r^{2}+z^{2}}}\ .
\label{Hamiltonian}%
\end{align}
Here $\psi=2\pi rA_{\phi}$ is the poloidal flux and is related to the magnetic
field by
\begin{equation}
\mathbf{B}=-\frac{1}{2\pi r}\frac{\partial\psi}{\partial z}\hat{r}%
-\frac{\partial A_{z}}{\partial r}\hat{\phi}+\frac{1}{2\pi r}\frac
{\partial\psi}{\partial r}\hat{z}, \label{B vect}%
\end{equation}
$V$ is the electrostatic potential, and we note that ideal MHD ultimately
comes from approximations based on Eq.\ref{Hamiltonian} and not
Eq.\ref{Kepler effective potential}. Because of axisymmetry the particle's
canonical angular momentum
\begin{equation}
P_{\phi}=m_{\sigma}r^{2}\frac{d\phi}{dt}+\ \frac{q_{\sigma}}{2\pi}%
\psi(r,z,t)=L_{\sigma}+\frac{q_{\sigma}}{2\pi}\psi
(r,z,t)\ \label{can ang momentum}%
\end{equation}
is invariant \cite{Schmidt:1979,Lehnert:1971}. Equation \ref{Hamiltonian} is
equivalent to the equation of motion
\begin{equation}
m_{\sigma}\frac{d\mathbf{v}}{dt\ }\ =q_{\sigma}\left(  \mathbf{E}%
+\mathbf{v}\times\mathbf{B}\right)  +m_{\sigma}\mathbf{g}%
\ \label{single particle motion}%
\end{equation}
with $\mathbf{E}=-\nabla V-\partial\mathbf{A}/\partial t$.

We examine solutions to Eq.\ref{single particle motion} in the $z=0$ plane for
various charge to mass ratios, a uniform magnetic field $\mathbf{B}=B_{z}%
\hat{z},$ and two representative $V(r)$ profiles. In order to see the
connection between the sense of particle injection and the magnetic field
direction, the coordinate system definition we use here is such that positive
$\phi$ is determined by the direction of injection of the particle, i.e., the
particle always has initial positive $d\phi/dt$ by assumption. This definition
means that $B_{z}$ could be positive or negative since the direction of the
$z$ axis is determined by the sense of particle injection (i.e., we are
defining the $z$ axis so that the particle is always injected in the
counterclockwise direction) and not by the direction of $\mathbf{B}$. Because
$B_{z}$ can be positive or negative, $\omega_{c\sigma}$ will have the usual
signage for $B_{z}>0$ but will have opposite polarity from convention if
$B_{z}<0.$ To see whether or not particles make Kepler orbits we track
particles starting with the same kinetic energy and $L_{\sigma}\ $as a neutral
particle undergoing an elliptical Kepler orbit. $a$ is defined to be the
radius of the initial location and the $x$ axis lies in the direction from the
central object to this location.

$L_{\sigma}\ $ is conserved if $q_{\sigma}=0,$ but if $q_{\sigma}\neq0$ then
$P_{\phi}$ rather than $L_{\sigma}$ is the conserved quantity
\cite{Schmidt:1979,Lehnert:1971}$.$ $\ B_{\phi}=0$ will be assumed (to be
justified later). Time is normalized to the Kepler frequency of a neutral
particle undergoing circular motion at $r=a$, i.e., to $\Omega_{0}%
=\sqrt{MG/a^{3}}$ and distances are normalized to $a.$\ The dimensionless
variables are then $\bar{r}=r/a,$ $\tau=\Omega_{0}t,$ $\mathbf{\bar{v}%
}=\mathbf{v}/\Omega_{0}a,$ $\bar{L}=L_{\sigma}/m_{\sigma}a^{2}\Omega_{0},$
$\bar{H}=H/m\Omega_{0}^{2}a^{2}$, $\ \bar{V}(\bar{r})=q_{\sigma}%
V(r)/m_{\sigma}\Omega_{0}^{2}a^{2},$ and using $\psi=B_{z}\pi r^{2},$
\begin{equation}
\bar{P}_{\phi}=\frac{P_{\phi}}{m_{\sigma}a^{2}\Omega}=\left(  \frac{d\phi
}{d\tau}+\frac{\omega_{c\sigma}}{\ 2\Omega_{0}}\right)  \ \bar{r}^{2}.
\label{Pphi norm}%
\end{equation}
The dimensionless $\bar{z}=0$ plane Hamiltonian is thus%
\begin{equation}
\bar{H}=\frac{1}{2}\bar{v}_{r}^{2}+\frac{1}{2}\left(  \frac{\bar{P}_{\phi}%
}{\bar{r}}-\dfrac{\omega_{c\sigma}}{2\Omega_{0}\ }\bar{r}\right)  ^{2}+\bar
{V}(\bar{r})-\frac{1}{\bar{r}}. \label{norm H}%
\end{equation}

Rotation of a plasma in a magnetic field polarizes the plasma radially and the
resulting $\bar{V}(\bar{r})$ corresponds to the voltage to which the plasma
capacitor is charged \cite{Lehnert:1971}. There is thus no natural $\bar
{V}(\bar{r})$ profile and hence no natural rotational velocity (e.g., see
Refs.\cite{Lehnert:1971,Ghosh:2004}). We consider two representative cases,
namely (i)\ $\bar{V}(r)=0$ and (ii) $\bar{V}(r)=2\bar{r}^{1/2}\omega_{c\sigma
}/\Omega_{0}.$ Case (i) corresponds to Eq.\ref{vg} while case (ii) corresponds
to the `Kepler' equilibria assumed in Refs.
\cite{Lovelace:1976,BalbusHawley:1991}. A third possibility, not discussed
here (see Ref.\cite{Howard:1999} instead),\ sets $\bar{V}(\bar{r})$ to give a
rotation velocity equal to that of the central object (so-called
\textquotedblleft co-rotational velocity\textquotedblright).\qquad

We consider all possible values of $\omega_{c\sigma}/\Omega_{0},$ namely
$\left\vert \omega_{c\sigma}/\Omega_{0}\right\vert <<1,$ $\left\vert
\omega_{c\sigma}/\Omega_{0}\right\vert >>1,$ and $\left\vert \omega_{c\sigma
}/\Omega_{0}\right\vert =\mathcal{O}(1)\ $with $\omega_{c\sigma}/\Omega_{0}$
either positive or negative. $\left\vert \omega_{c\sigma}/\Omega
_{0}\right\vert >>1$ is typical for electrons, ions, and large charge to mass
ratio dust grains, whereas $\left\vert \omega_{c\sigma}/\Omega_{0}\right\vert
<<1$ corresponds to dust grains with very small charge to mass ratios or
macroscopic charged particles such as spacecraft \cite{Garrett:2000}. The
voltage $V_{d\text{ }}$ to which a dust grain becomes charged depends on the
charging mechanism and the dust grain size; $V_{d\text{ }}$ typically lies in
the range $1$ volt $<\left\vert V_{d}\right\vert <100$ volts. \thinspace Since
the dust grain charge is $Q_{d}=4\pi\varepsilon_{0}r_{d}V_{d}$, the charge to
mass ratio $Q_{d}/m_{d}=3\varepsilon_{0}V_{d}/r_{d}^{2}\rho_{d}$ lies in the
range $\ 10^{-4}-10^{2}$ C/kg for typical dust grain radii $\ 0.1$ $\mu$m
$<r_{d}<10$ $\mu$m and typical dust grain intrinsic mass density $\rho
_{d}=10^{3}$ kg m$^{-3}.$ The dust grain $\omega_{cd}/\Omega$ ratio is thus
9-15 orders of magnitude smaller than that of an electron and 6-12 orders of
magnitude smaller than that of an ion.

The last two terms in Eq.\ref{norm H} can be written as a normalized effective
potential
\begin{equation}
\bar{\chi}(\bar{r})=\frac{1}{2}\left(  \frac{\bar{P}_{\phi}}{\bar{r}}%
-\dfrac{\omega_{c\sigma}}{2\Omega_{0}\ }\bar{r}\right)  ^{2}+\bar{V}(\bar
{r})-\frac{1}{\bar{r}}\ . \label{chi}%
\end{equation}
If $q_{\sigma}=0$ then $\omega_{c\sigma}=0$, $\bar{V}=0$ and $\bar{\chi
}\rightarrow\bar{\chi}_{Kepler}$ in which case $\ $Kepler dynamics
\cite{Landau:1960} is retrieved.

However, when $\ $ $\omega_{c\sigma}\neq0$ $\ $and $\bar{V}(\bar{r})$ is
arbitrary, the dynamics is non-Keplerian and Eq. \ref{chi} has minima when%
\begin{equation}
\ \left(  \frac{\bar{P}_{\phi}}{\bar{r}^{2}}\right)  ^{2}-\left(
\dfrac{\omega_{c\sigma}}{2\Omega_{0}\ }\right)  ^{2}\ -\frac{1}{\bar{r}}%
\frac{\partial\bar{V}}{\partial\bar{r}}-\frac{1}{\bar{r}^{3}}%
=0\ .\label{min chi}%
\end{equation}
By using Eq.\ref{Pphi norm}, Eq.\ref{min chi} can be recast as
\begin{equation}
\ \ \left(  \frac{d\phi}{d\tau}\right)  ^{2}+\ \frac{\omega_{c\sigma}%
}{\ \Omega_{0}}\frac{d\phi}{d\tau}\ -\frac{1}{\bar{r}}\frac{\partial\bar{V}%
}{\partial\bar{r}}-\frac{1}{\bar{r}^{3}}=0\ \label{quad 0}%
\end{equation}
so a particle with $\bar{H}$ equal to the effective potential minimum has an
angular velocity%
\begin{equation}
\frac{d\phi}{d\tau}\ =-\frac{\omega_{c\sigma}}{\ 2\Omega_{0}\ }\pm
\sqrt{\left(  \frac{\omega_{c\sigma}}{2\ \Omega_{0}}\right)  ^{2}+\frac
{1}{\bar{r}}\frac{\partial\bar{V}}{\partial\bar{r}}+\frac{1}{\bar{r}^{3}}%
}.\label{quadratic}%
\end{equation}

\bigskip

Figure 1 plots $\chi(\bar{r})$ and numerically calculated $x$-$y$ plane
trajectories for a range of $\omega_{c\sigma}/\Omega_{0}$ values and for the
two $\bar{V}(\bar{r})$ cases. In all trajectory calculations the particle
initial position is $\bar{x}=1,\bar{y}=0$ and the initial velocity is $\bar
{v}_{x}=0.4,$ $\bar{v}_{y}=1\ $(i.e., particles start at the same position
with the same initial velocity and the same initial mechanical angular
momentum $L_{\sigma}$). The trajectories are calculated from $\tau=0$ to
$4\pi$ (i.e., two circular Kepler orbit periods) and the energy $\bar{H}$ is
shown as a dashed line in the effective potential plots. An $\bar{r}=1$
reference circle (dashed) \ is\ shown in the trajectory plots.

Figure 1 shows that the trajectory depends strongly on both the $\bar{V}%
(\bar{r})$ profile and on $\omega_{c\sigma}/\Omega_{0}.$ The $\omega_{c\sigma
}/\Omega_{0}=0$ situation (fifth row of Fig. 1) is a classic elliptical Kepler
orbit as prescribed by Eq.\ref{Kepler effective potential} and is independent
of $\bar{V}(\bar{r})$ because a neutral particle is insensitive to
electromagnetic fields. However, when $\omega_{c\sigma}/\Omega_{0}$ is finite,
Fig. 1 shows that the effective potential and trajectories differ
qualitatively from the classic neutral particle effective potential and
elliptical Kepler orbit. It is therefore incorrect to characterize a plasma
composed of particles with $\left\vert \omega_{c\sigma}/\Omega_{0}\right\vert
>>1$ as being in a Kepler orbit (as done in the MRI\ and UID models) because
particles in such a plasma are not governed by
Eq.\ref{Kepler effective potential} and, for example, do not make elliptical
orbits with the central object at one focus of the ellipse (such orbits are
the \textquotedblleft hallmark\textquotedblright\ of Kepler dynamics).

Insight into the orbits shown in Fig. 1 can be obtained by examining solutions
to Eq.\ref{quadratic}. For $\left\vert \omega_{c\sigma}/2\Omega_{0}\right\vert
<<1$\ and $\bar{V}(\bar{r})=0,$ Eq.\ref{quadratic} has the roots
\begin{equation}
\frac{d\phi}{d\tau}\ =\pm\frac{1}{\bar{r}^{3/2}}-\frac{\omega_{c\sigma}%
}{\ 2\Omega_{0}\ }\ \label{Kepler}%
\end{equation}
so heavy charged dust grains with $\bar{H}\ $ equal to the minimum of the
effective potential make circular orbits with a small $\omega_{c\sigma
}/2\Omega_{0}$ correction to the Kepler frequency. Heavy charged dust grains
with $\bar{H}$ slightly above this minimum will make precessing elliptical
Kepler orbits having small $\omega_{c\sigma}/2\Omega_{0}$ corrections \ (see
$\omega_{c\sigma}/\Omega_{0}=\pm0.1$ cases in Fig. 1) and these corrections
will increase with the charge to mass ratio.

For $\left\vert \omega_{c\sigma}/2\Omega_{0}\right\vert >>1$ and $\bar{V}%
(\bar{r})=0$ the two roots of Eq.\ref{quadratic} are
\begin{equation}
\frac{d\phi}{d\tau}\ =-\frac{\omega_{c\sigma}}{\ \Omega_{0}\ }\text{ , }%
\frac{d\phi}{d\tau}\ =\ \frac{1}{\bar{r}^{3}}\frac{\Omega_{0}}{\omega
_{c\sigma}}. \label{drift solution}%
\end{equation}
The first root corresponds\ to a so-called axis-encircling cyclotron orbit
\cite{Schmidt:1979}; this root is not likely to be physically realizable in
astrophysical situations since its corresponding azimuthal velocity exceeds
the Kepler velocity by the large ratio $\left\vert \omega_{c\sigma}/\Omega
_{0}\right\vert $. Normal cyclotron orbits correspond to a particle
oscillating \cite{Schmidt:1979,Lehnert:1971} in $r$ about a local minimum of
$\chi(r)$ and are associated with the second root in Eq.\ref{drift solution}.
The second root is just $\mathbf{v}_{g}$ prescribed by Eq.\ref{vg}
\cite{Alfven:1950,Spitzer:1956,Sturrock:1994} and, as discussed above, is
\textit{smaller} than $\mathbf{v}_{K}$ by the factor $\left\vert \Omega
_{0}/\omega_{c\sigma}\right\vert $. In reality $\left\vert \Omega_{0}%
/\omega_{c\sigma}\right\vert $ would be so enormous that electrons and ions
would have negligible azimuthal displacement during one Kepler period of a
neutral particle. These slow drift orbits are shown in the top and bottom rows
of case (i) in Fig.1. In accordance with Eq.\ref{vg} heavy particles drift
faster, negative and positive particles drift in opposite directions, and the
drift velocity decreases as $B_{z}$ increases.

For case (ii), $\bar{V}(\bar{r})=\ 2\bar{r}^{1/2}\omega_{c\sigma}/\Omega$, and
so Eq.\ref{quad 0} becomes%
\begin{equation}
\left(  \frac{d\phi}{d\tau}-\frac{1}{\bar{r}^{3/2}}\right)  \left(
\frac{d\phi}{d\tau}^{\ }+\frac{1}{\bar{r}^{3/2}}+\frac{\omega_{c\sigma}%
}{\Omega_{0}}\right)  \ \ =0 \label{Kepler phitdot quad}%
\end{equation}
where one root is the circular Kepler-like orbit $d\phi/d\tau=1$ with $\bar
{r}=1.$ Although the $d\phi/d\tau^{\ }=1$ root looks superficially like a
neutral particle Kepler orbit, the orbits are not elliptical, but nearly
circular and, as in tokamaks and spheromaks, stay within a poloidal Larmor
orbit of a constant $\psi$ surface \cite{Lehnert:1971,Bellan:2000}. The
effective potential minimum has the same radial location as
Eq.\ref{Kepler effective potential} but the profile is an extremely narrow
trough with a large vertical offset (positive or negative, depending on the
charge polarity), not a shallow broad well as for
Eq.\ref{Kepler effective potential}.

\section{Orbits of particles with zero canonical angular momentum: dynamo for
driving astrophysical jets}

A strange behavior is evident in the $\omega_{c\sigma}/\Omega_{0}=-2.0$ row of
Fig. 1: the effective potential goes to minus infinity on the left and the
particle spirals inwards toward the origin in the $x$-$y$ trajectory plots.
This corresponds to $P_{\phi}=0\ $and is unlikely for electrons or ions
because they typically have $\left\vert \omega_{c\sigma}/\Omega_{0}\right\vert
>>1.$ However, $P_{\phi}=0$ could occur for dust grains because, being heavy,
dust grains have $\left\vert \omega_{c\sigma}/2\Omega_{0}\right\vert $ many
orders of magnitude smaller than electrons or ions. As seen from
Eq.\ref{Pphi norm},\ $\bar{P}_{\phi}=0$ occurs if $\ d\phi/d\tau
=-\omega_{c\sigma}/2\Omega_{0}$ or, in un-normalized quantities $P_{\phi}=0$
occurs when $d\phi/dt=-\omega_{c\sigma}/2$ in which case $d\phi/dt$ also
becomes invariant. In this situation the normalized effective potential,
Eq.\ref{chi}, reduces to
\begin{equation}
\bar{\chi}(\bar{r})=\frac{1}{8\ }\left(  \ \dfrac{\omega_{c\sigma}}%
{\ \Omega_{0}\ }\right)  ^{2}\bar{r}^{2}+\bar{V}(\bar{r})-\frac{1}{\bar{r}%
}\ \ \text{\ } \label{special effective potential}%
\end{equation}
which has the remarkable feature that no centrifugal repulsion exists so the
particle always falls towards $\bar{r}=0$ with constant $d\phi/d\tau$ (i.e.,
it spirals in). This differs qualitatively from $\bar{P}_{\phi}\neq0$
particles which are constrained to orbit at a fixed average radius. The
normalized radial force acting on a $\bar{P}_{\phi}=0$ particle is
\begin{equation}
\bar{F}=-\frac{\partial\bar{\chi}}{\partial\bar{r}}=-\frac{1}{4\ }\left(
\ \dfrac{\omega_{c\sigma}}{\ \Omega_{0}\ }\right)  ^{2}\bar{r}-\frac
{\partial\bar{V}}{\partial\bar{r}}-\frac{1}{\bar{r}^{2}}\ \text{\ }
\label{special force}%
\end{equation}
which is negative for any potential having $\partial\bar{V}/\partial\bar
{r}\geq0$. Thus $\bar{F}$ is negative for both the $\bar{V}=0$ and the
`Kepler' potential $\bar{V}(\bar{r})=\ 2\bar{r}^{1/2}\omega_{c\sigma}/\Omega.$
If there is a distribution of dust grain sizes, then a corresponding
distribution of $\omega_{c\sigma}/\Omega_{0}$ values will develop and some
subset will have $\omega_{c\sigma}/\Omega_{0}=-2$. The situation
$\omega_{c\sigma}/\Omega_{0}=-2$ could occur for negatively charged dust
grains injected with $B_{z}>0$ or for positively charged dust grains with
$B_{z}<0.$ This latter case is the likely one and if one were to adopt the
common convention that the $z$ axis is defined by the direction of
$\mathbf{B\,}$, this would correspond to retrograde injection.

Dust grains accreting to a circumstellar disk will typically absorb stellar
UV\ photons \cite{Wolf:2003} and become positively charged by emitting
photo-electrons \cite{Verheest:2000,Sickafoose:2000}. Some of these dust
grains will satisfy $d\phi/dt=-\omega_{c\sigma}/2$ if $B_{z}<0$ and so have
$\bar{P}_{\phi}=P_{\phi}=0.$ These dust grains have mechanical angular
momentum $L_{\sigma}=m_{\sigma}r^{2}d\phi/dt$ at the instant before becoming
charged by photo-emitting electrons, i.e., they have mechanical angular
momentum $L_{\sigma}=-q_{\sigma}\psi/2\pi$ at the instant before they become
charged. Since neither $r$ nor $d\phi/dt$ is changed at the instant of
charging, their canonical angular momentum $P_{\phi}=L_{\sigma}+q\psi/2\pi$
becomes zero at the instant after charging.

The infalling $\bar{P}_{\phi}=0$ dust grains will accumulate at small $\bar
{r}\ $and create a positive space charge there. The photo-emitted electrons,
stranded at large $\bar{r}$ (since electrons have $P_{\phi}\neq0$), will
create a corresponding negative space charge at large $\bar{r}.$ The positive
and negative space charges will tend to cancel any polarization charge, e.g.,
the polarization charge associated with an initial $\bar{V}(\bar{r}%
)=\ 2\bar{r}^{1/2}\omega_{c\sigma}/\Omega$ potential. Accumulation of
infalling $\bar{P}_{\phi}=0$ positive dust grains will eventually create an
outward radial electric field $E_{r}^{\ast}$ (i.e., opposite direction to that
associated with the $\bar{V}(\bar{r})=\ 2\bar{r}^{1/2}\omega_{c\sigma}/\Omega$
potential). This accumulation $\ $ will cease when $E_{r}^{\ast}$ becomes
sufficiently large to create a force $q_{\sigma}E_{r}^{\ast}$ which cancels
$F.$ In un-normalized variables this cancellation occurs when $E_{r}^{\ast
}=\ \partial V_{other}/\partial r+\left(  r\omega_{c\sigma}^{2}/4\ +MG/r^{2}%
\ \right)  m_{\sigma}/q_{\sigma}$ where $V_{other}$ is the potential profile
that would exist due to particles other than the accumulating $P_{\phi}=0$
dust grains. The inward falling $P_{\phi}=0$ dust grains constitute a radially
inward conduction current so $\mathbf{J\cdot E}$ is negative and the system
converts gravitational potential energy into available electrical power; i.e.,
it is a dynamo. The positive voltage near $r=0$ will drive bipolar axial
electric \ currents $I$ outwards from the $z=0$ plane. These currents will
deplete the positive space charge which will result in a net force
$F-q_{d}E_{r}^{\ast}$ that will drive additional $\bar{P}_{\phi}=0$ dust
grains towards $r=0$ where they will replenish the positive space charge.
Thus, the system continuously converts the gravitational potential energy of
the $P_{\phi}=0$ accreting dust grains into a battery-like electrostatic
potential which drives the poloidal current of an astrophysical jet. The jet
itself is accelerated by the $\partial B_{\phi}^{2}/\partial z$ force
\cite{Bellan:1992,Bellan:2003,You:2005} associated with the axial
non-uniformity of the jet poloidal current $I(r,z)$ since $B_{\phi}%
(r,z)=\mu_{0}I(r,z)/2\pi r.$ The anti-symmetry of $I$ with respect to $z$
means that $B_{\phi}=0$ at $z=0$ consistent with the assumption made earlier.

We note that the ability of $P_{\phi}=0$ particles to cross magnetic flux
surfaces has recently been observed in a laboratory experiment
\cite{Tripathi:2007}.

The assumption used in this paper that $B_{z}$ is spatially uniform is a
simplifying idealization that enables the analysis to be both brief and
focussed on the distinction between Kepler and charged particle orbits.
However, an actual accretion disk will almost certainly have $B_{z}$ depend on
both $r$ and on $z$ so in order to satisfy $\nabla\cdot\mathbf{B}=0$ there
will also have to be a $B_{r}(r,z).$ This indicates that the axisymmetric
magnetic field would be best described using a poloidal flux function
$\psi(r,z)$ i.e., $\mathbf{B}(r,z)=(2\pi)^{-1}\left(  \nabla\psi\times
\nabla\phi+\mu_{0}I(r,z)\nabla\phi\right)  .$ This more general description of
the magnetic field has been used in Ref.\cite{Bellan:2007}, a much lengthier
analysis, where three dimensional particle orbits in an approximately
self-consistent magnetic field are considered using a generalization of the
Hamiltonian method presented here. Specifically, the poloidal flux function
$\psi(r,z)$ in Ref.\cite{Bellan:2007} results from a toroidal current due to
toroidal motion of charged particles and particles are not restricted to the
$z=0$ plane as in the present paper.

The dust grains might be so densely packed as to be optically thick in which
case photons from the central object would not reach the dust and the dust
would not become charged; this would constitute a so-called `dead-zone'. The
condition for a dust cloud to be optically thick is $n\sigma L>1$ where $n$ is
the dust density, $\sigma$ is the dust cross-section, and $L$ is the
characteristic length scale of the dust cloud. However, the condition for a
dust cloud to be collisional is also $n\sigma L>1$ and so dust grains in an
optically thick cloud would be collisional. This collisional, optically thick
state would likely be transient, because collisions are expected to cause
coagulation of the dust grains \cite{Weidenschilling:2000} in which case their
radius $r_{d}$ will increase. Since the mass of an individual dust grain is
$m_{d}=4\pi\rho_{d}r_{d}^{3}/3$ and since coagulation does not change the
total mass $M$ of all the dust grains, the number $N=M/m$ of dust grains and
hence the density $n\sim N/L^{3}$ of dust grains scales as $r_{d}^{-3}.$
Because the dust grain cross-section $\sigma$ scales as $r_{d}^{2}$, the
product $n\sigma\ $scales as $r_{d}^{-1}\ $and so $n\sigma$ decreases as a
result of coagulation. Coagulation of dust grains will thus reduce $n\sigma$
until $n\sigma L$ becomes less than unity in which case the dust cloud will
become collisionless and optically thin. At this point dust would become
charged (ionized) via photo emission and commence the collisionless
trajectories discussed here. The dead zones would thus disappear as a result
of coagulation. This issue is discussed in more detail in
Ref.\cite{Bellan:2007}.

Ref.\cite{Bellan:2007} discusses several other issues including:\ the charging
rate of dust grains (i.e., effective ionization rate of dust grains),
collisions of dust grains with gas and other dust grains, and the topological
properties of astrophysical jets. These various issues are used to define a
parameter space for astrophysical jets powered by the gravitational
energy\ released by accreting $P_{\phi}=0$ dust grains$.$ A self-consistent
set of parameters is given for the example of the circumstellar accretion disk
of a young stellar object (YSO).

\bigskip

\bigskip\pagebreak

%

%TCIMACRO{\FRAME{fhFU}{5.546in}{8.3048in}{0pt}{\Qcb{Effective potentials and
%$\bar{x}$-$\bar{y}$ plane trajectories for particles starting at $\bar
%{x}=1,\bar{y}=0$ with initial velocity $\bar{v}_{x}=0.4$, $\bar{v}_{y}=1$ with
%range of $\omega_{c\sigma}/\Omega_{0}$ values and two $\bar{V}(\bar{r})$
%cases. Elliptical Kepler orbits (i.e., Eq.\ref{Kepler effective potential}
%effective potential) occur only when $\omega_{c\sigma}/\Omega=0.$ Particles
%with $\omega_{c\sigma}/\Omega_{0}=-2$ fall towards $\bar{r}=0$ and have
%$\bar{P}_{\phi}=0.$ No particles make Kepler-like elliptical orbits when
%$\left\vert \omega_{c\sigma}/\Omega_{0}\right\vert >>1.$}}{}%
%{effective-potential-orbits.eps}{\special{ language "Scientific Word";
%type "GRAPHIC";  maintain-aspect-ratio TRUE;  display "USEDEF";
%valid_file "F";  width 5.546in;  height 8.3048in;  depth 0pt;
%original-width 6.4913in;  original-height 9.7369in;  cropleft "0";
%filename 'EFFECTIVE-POTENTIAL-ORBITS.eps';file-properties "XNPEU";}} }%
%BeginExpansion
\begin{figure}
[h]
\begin{center}
\includegraphics[
height=8.3048in,
width=5.546in
]%
{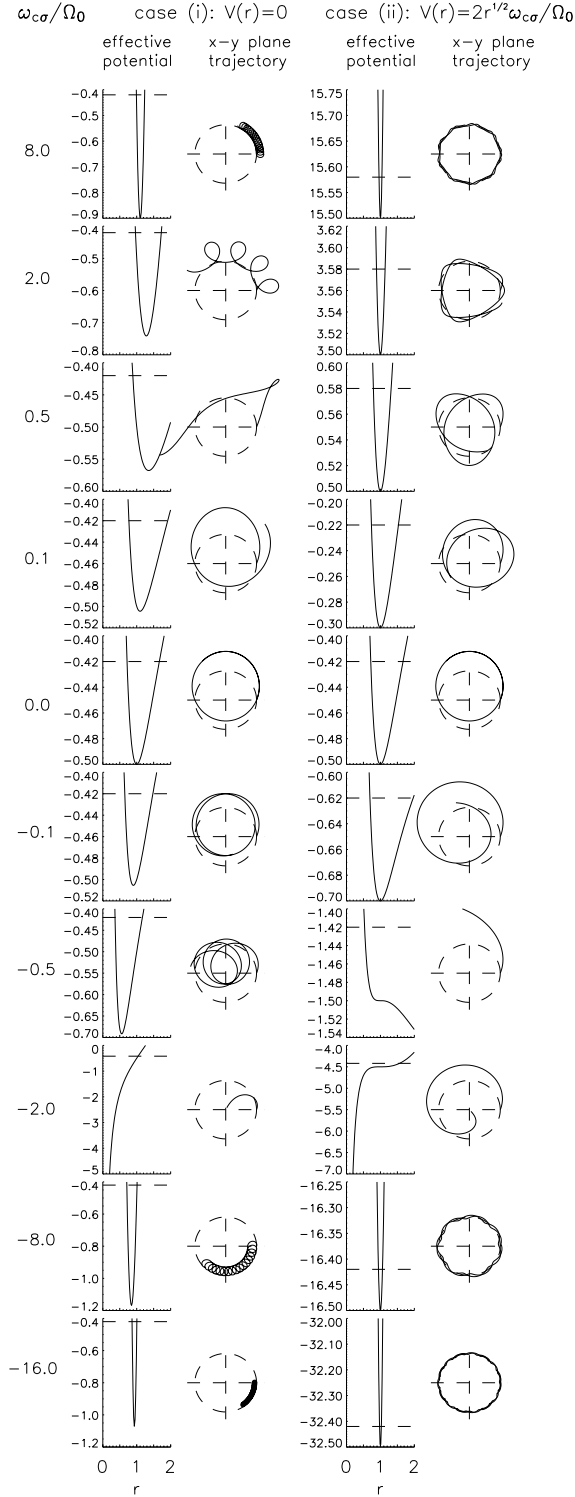}%
\caption{Effective potentials and $\bar{x}$-$\bar{y}$ plane trajectories for
particles starting at $\bar{x}=1,\bar{y}=0$ with initial velocity $\bar{v}%
_{x}=0.4$, $\bar{v}_{y}=1$ with range of $\omega_{c\sigma}/\Omega_{0}$ values
and two $\bar{V}(\bar{r})$ cases. Elliptical Kepler orbits (i.e.,
Eq.\ref{Kepler effective potential} effective potential) occur only when
$\omega_{c\sigma}/\Omega=0.$ Particles with $\omega_{c\sigma}/\Omega_{0}=-2$
fall towards $\bar{r}=0$ and have $\bar{P}_{\phi}=0.$ No particles make
Kepler-like elliptical orbits when $\left\vert \omega_{c\sigma}/\Omega
_{0}\right\vert >>1.$}%
\end{center}
\end{figure}
%EndExpansion

\end{document}